\documentclass[referee]{raa}            

\usepackage{graphicx,times}             
\usepackage{natbib}
\usepackage{orcidlink}
\usepackage{fontspec}
\usepackage{hyperref}
\usepackage{amssymb,amsmath} 
\bibpunct{(}{)}{;}{a}{}{,}

\begin{document}

\title{Accuracy, Convergence and Consistency of Dark Matter Evolution with the PhotoNs-3 Code}

   \volnopage{Vol.0 (202x) No.0, 000--000}      
   \setcounter{page}{1}          

   \author{Xiaoyang Zhang 
      \inst{1,2}\orcidlink{0009-0009-7226-0040}
   \and Ming Li
      \inst{1,2}\orcidlink{0000-0002-1318-4828}
   \and Qiao Wang
      \inst{1,2,*}\footnotetext{$*$Corresponding author:{\it qwang@nao.cas.cn}}\orcidlink{0000-0003-2153-7758}
   \and Liang Gao
      \inst{3,4}\orcidlink{0009-0006-3885-9728}
   }

   \institute{National Astronomical Observatories, Chinese Academy of Sciences, Beijing 100101, China \\
        \and
    School of Astronomy and Space Science, University of Chinese Academy of Sciences, Beijing 100048, China \\
        \and
    Institute for Frontiers in Astronomy and Astrophysics, Beijing Normal University, Beijing 102206, China \\
        \and
    School of Physics and Microelectronics, Zhengzhou University, Zhengzhou 450001, China \\
        \vs\no
   {\small Received 202x month day; accepted 202x month day}}

\abstract{
We systematically investigate the accuracy and convergence of PhotoNs-3, a cosmological N-body code employed for the HyperMillennium simulation. In particular, we examine its convergence with respect to parameters that affect the accuracy of the force calculation and the time evolution, such as the softening length and the adaptive time-stepping scheme, and find that it converges very well. Furthermore, using GADGET as the benchmark, we design four sets of simulations to compare the mutual consistency. We compare the power spectrum of the density field, halo density profiles, and the halo and subhalo mass functions. The difference between PhotoNs-3 and the Gadget codes is at the same level as that between Gadget-3 and Gadget-4. Meanwhile, PhotoNs-3 generally achieves a significant computational speedup. This work establishes that the accuracy and reliability of PhotoNs-3 and its simulations are sufficient for Stage IV galaxy surveys.
}

   \authorrunning{Zhang X. et al.}            
   \titlerunning{Accuracy, Convergence and Consistency of PhotoNs-3} 

   \maketitle

%
%

\section{Introduction}

Current and forthcoming wide-field surveys will map the Universe with unprecedented statistical power, enabling precise constrain of cosmological models \citep{2012arXiv1208.4012G, 2019ApJ...873..111I, DESI2022, 2024ApJ...973L..14D, 2025A&A...697A...1E, 2026SCPMA..6939501C}. Interpreting these measurements requires theoretical predictions whose accuracy matches the observational and modeling systematics. On large scales, linear perturbation theory suffices, whereas nonlinear structure formation of Dark Matter (DM) on small scales must be modeled with N-body simulations \citep{Heitmann}. Since these simulations are used to calibrate clustering statistics and halo abundances, numerical errors in N-body predictions propagate directly into the systematic error budget of survey analyses. It is therefore essential to quantify the accuracy, convergence, and cross-code consistency of codes used for extreme-scale simulations in precision cosmology studies \citep{2021MNRAS.506.4210I, Maksimova_2021, 2022ApJS..259...15F, 2021ApJS..252...19H, 2023MNRAS.524.2556H, 2025SCPMA..6809511H, 2025A&A...697A...5E, hyper26}.

However, reaching this accuracy over the volumes and resolutions demanded by current and forthcoming surveys is computationally challenging: particle counts become extremely large, while memory capacity, communication volume, and load imbalance increasingly constrain conventional CPU-only Tree and Particle Mesh (PM) implementations. These challenges motivate GPU-accelerated, memory-efficient N-body code. Among codes developed for such extreme-scale simulations, PhotoNs has been designed specifically for cosmological simulations on heterogeneous supercomputers \citep{photons}. It implements an algorithm of Particle-Mesh Fast-Multipole-Method (PM-FMM). The third version of the code, PhotoNs-3, revises the parallelization strategy, GPU acceleration, and memory management to mitigate communication and memory bottlenecks that emerge when simulations reach extreme scales \citep{SSPMA26}, such as the Ultramarine and HyperMillennium (HM) simulations \citep{ultra22, hyper26}.

Various simulation codes employ N-body method to solve gravity for cosmological structure formation, including GADGET \citep{GADGET2, Gadget4}, RAMSES \citep{RAMSES02}, PKDGRAV \citep{PKDGRAV, PKDGRAV3}, SWIFT \citep{swift16}, ABACUS \citep{ABACUS}, CUBEP$^3$M \citep{2013MNRAS.436..540H}, GreeM \citep{2021MNRAS.506.4210I}, HACC \citep{2016NewA...42...49H} and CUBE2 \citep{2026SCPMA..6969511Y}. Comparisons between independently developed codes provide a fundamental tool for validating simulation results, since such comparisons expose numerical systematic effects that no single code can identify on its own. Several multi-code comparisons have been performed for various statistics, such as clustering statistics, halo abundances, and internal halo properties \citep{Sch, Garrion19, DESI22, Bayer, Euclidc}. In such studies, GADGET is one of the most popular benchmarks.

In this work, we examine the accuracy and convergence of PhotoNs-3, assess the new numerical methods and optimization introduced in the recent release, and thereby demonstrate the reliability of the HM simulation. The rest of the paper is structured as follows. In Section~2, we briefly introduce PhotoNs-3 and present its convergence tests. In Section~3, we compare our simulation results with GADGET in several aspects and  computational performance. Finally, we summarize our conclusions in Section~4.

\section{upgrades, Accuracy \& Convergence}

\textsc{PhotoNs} is an \(N\)-body simulation code for formation and evolution of cosmological large-scale structure. Since the beginning, it has been designed as high-performance software for large-scale, heterogeneous supercomputing platforms: early versions adopted the TreePM method with octree structure \citep{photons}; starting from the second version, the PM-FMM method was introduced, together with an accelerated implementation on NVIDIA GPUs, whose accuracy was also validated \citep{FMM, photonsGPU}. The third version further optimizes the implementation of the multipole expansion; more importantly, to adapt to next-generation supercomputing platforms, its large-scale parallelization strategy has been thoroughly rewritten using a two-dimensional pencil decomposition, so that the spatial layout of particle data better matches the global MPI communication and memory-access patterns of the large-scale FFTs required by the convolution in the PM solver \citep{ultra22}. In addition, the computing kernels have been extended to HIP-based GPU accelerators (GPGPU) in ORISE/Sugon-8000 supercomputer and SW26010-pro CPUs in OceanLight supercomputer. For large-scale runs, PhotoNs-3 also incorporates memory optimizations and employs a quantization scheme to compress data and reduce the memory bottleneck \citep{SSPMA26}. These advances are summarized in Table~\ref{tab:codeversion}. Moreover, many technical details and optimization strategies that have not been systematically documented in the literature jointly enable trillion-particle-scale cosmological simulations, while potentially introducing uncertainties and biases. This work aims to examine whether, after these advances and extensive optimizations—particularly in the gravity solver and time-stepping scheme—\textsc{PhotoNs}-3 still achieves the accuracy and convergence required by the scientific goals.

\begin{table}[htpb]
    \centering
    \begin{tabular}{cccccc}
    \hline
    \hline
    version & gravity solver & tree structure & domain & heterogeneity & optimization \\
    \hline
    1 & TreePM & octree & filling curve & many core & kernel \\
    2 & PM-FMM & k-d tree & k-d tree & GPU & task, kernel \\
    3 & PM-FMM & k-d tree & pencil shape & GPGPU, SW26010 & memory, kernel  \\
    \hline
    \end{tabular}
    \caption{Summary of the three versions of the PhotoNs code.}
    \label{tab:codeversion}
\end{table}

Gravity is the dominant force in cosmological structure formation, and computing the gravitational field is essential in any N-body simulation. The PM-FMM in PhotoNs is designed with heterogeneous platforms. Its key idea is to decompose the gravitational potential: $\varphi = \varphi_{\rm long} + \varphi_{\rm short}$, based on the scale radius of $r_{\rm s}$, so that distant interactions are decoupled and the massive communication is required in large-scale simulations. The short-range component is computed with a compensated force kernel rather than an original Newtonian force. The short-range acceleration is written as
\begin{equation} 
\label{eq:short}
\mathbf{g}_{\rm short}(\mathbf{r}) = -\frac{GM \mathbf{r}}{r^3}\, T\!\left(\frac{r}{2r_s}\right),
\end{equation}
where the shape function $T(x)$ reads
\begin{equation} 
\label{eq:shapefunc}
T(x) = {\rm erfc}(x) + \frac{2x}{\sqrt{\pi}}\, e^{-x^2},
\end{equation}
with ${\rm erfc}(x)$ the complementary error function. The interaction is truncated beyond the cutoff radius $R_{\rm cut} = 6 \Delta_g$ ($\Delta_g$ is the mesh width in PM subroutine), consistent with previous simulations.

The short-range gravity solver involves two key operators: the multipole-to-local translation (M2L) and the particle-to-particle interaction (P2P). The tree traversal is governed by the Multipole Acceptance Criterion (MAC) with an opening angle $\theta$, and the interactions at the leaf nodes are evaluated by direct pairwise summation, with the maximum number of particles per leaf capped at $N^{\rm max}_{\rm leaf}$.

Evaluating the short-range interaction directly is computationally expensive. In the current implementation, we accelerate it by performing P2P interactions at the leaf nodes of the tree, instead of the conventional tree-node interactions. This choice, however, leads to a different accuracy and convergence behavior. The parameters and factors that control the accuracy of the PM-FMM under this scheme are summarized:

\begin{itemize}
\item[1.] \textbf{PM-FMM parameters:} The maximum leaf size and the opening angle, denoted as $N^{\rm max}_{\rm leaf}$ and $\theta$, respectively, are discussed in the algorithm paper \citep{FMM}. In HM run and this work, we employ $N^{\rm max}_{\rm leaf} = 256$, $\theta = 0.4$ and the cutoff radius $R_{\rm cut} = 4.5\,r_s$. PM method with grid size $N_{\rm mesh} = N_{\rm part}$, where $N_{\rm part}$ is particle number on one side.

\item[2.] \textbf{Numerical implementation:} The round-off precision, especially from heterogeneous optimization \citep{photonsGPU}, and the coordinate quantization scheme are designed to match the double precision of GADGET, as discussed in the technical review \citep{SSPMA26}.

\item[3.] \textbf{Softening scheme:} The softening length, denoted as $\varepsilon_{\rm soft}$, presents the geometric scale of mass points and the spatial resolution limit of the simulation, preventing numerical divergence in gravitational interactions ($\varepsilon_{\rm soft} = 3\,h^{-1}\,{\rm kpc}$ in the HM run). Its functional shape has been modified for the GPU optimization, we examine its effect in Section 2.1.

\item[4.] \textbf{Time stepping:} We also adopt an adaptive time-stepping scheme with power-of-two subdivisions. In high-precision numerical simulations, very fine time steps are usually required; in the HM simulation, more than 17,000 steps were used. We study it in Section 2.2.

\item[5.] \textbf{Configuration:} The resolution and cosmological parameters can affect the simulation results; we investigate them through cross-validation against GADGET. See Section 3 for details.
\end{itemize}

The simulations in this work are carried out on two platforms. The first is a GPU-accelerated machine equipped with two NVIDIA A100 GPUs, each with 80 GB of memory, connected to the host via PCIe. The host comprises dual 16-core Intel Xeon Gold 6326 CPUs with 2.9 GHz and 1.5 TB of main memory. The second platform is the ORISE supercomputer, whose computing nodes are each equipped with dual 8-core x86 CPUs, 128 GB of host memory, and four HIP-based graphics processing units (GPUs).

In this section, we present the convergence tests for the softening length and the adaptive time-stepping scheme, using the fiducial setup of the HM simulation. All tests are performed with the same code version adopted in the HM simulation project, namely PhotoNs-3.7. Specifically, we adopt the cosmological parameters of the HM simulation from Planck 2018 \citep{Planck2018}: $\Omega_M = 0.3111$, $\Omega_\Lambda = 0.6889$, $H_0 = 67.66\,{\rm km}\,{\rm s}^{-1}\, {\rm Mpc}^{-1}$, $\sigma_8 = 0.8102$, and $n_s = 0.9665$. The HM simulation box has a comoving size of $L_{\rm box} = 2.5\,h^{-1}\,{\rm Gpc}$ on a side, with particle mass of $3.21 \times 10^8~h^{-1}\,{\rm M}_\odot$, $N_{\rm mesh} = 16128$, and a softening length of $\varepsilon_{\rm soft} = 3\,h^{-1}\,{\rm kpc}$ \citep{hyper26}.

\subsection{Softening scheme}
\label{sec:soft}

In PhotoNs-3, the softening scheme has been revised and the softened potential now is induced by a flatten core of density profile instead of the Plummer shape, so that the gravitational acceleration $\mathbf{g}(\mathbf{r})$ reads
\begin{equation}
\label{soft-eq}
\mathbf{g}(\mathbf{r}) =
\begin{cases}
-GM\mathbf{r}/h^3 & (|\mathbf{r}| \leq h), \\[2mm]
-GM\mathbf{r}/|\mathbf{r}|^3 & (|\mathbf{r}| > h),
\end{cases}
\end{equation}
where $h = 3 \varepsilon_{\rm soft}/2$ is the critical radius of the kernel
\footnote{The gravitational constant $G$ adopted in the values of CODATA 2010 differs slightly from that in the earlier codes, leading to systematic deviations at the $10^{-4}$ level. In this paper, we adopt a consistent gravitational constant throughout.}. 
To examine the convergence of this scheme, we perform a series of tests in which the softening length is varied. The softening length is expressed in units of the mean interparticle separation, $\varepsilon_{\rm soft} = \xi ( L_{\rm box}/N_{\rm part} ) $ and $\varepsilon_{\rm soft} = \xi \Delta_g$ in the condition of $N_{\rm mesh} = N_{\rm part}$. The conventional choice is $\xi = 0.02$. The optimal value of the softening length is discussed in \citet{Soft03, Soft19}.

\begin{figure}[htpb]
    \centering
    \includegraphics[width=0.95\linewidth]{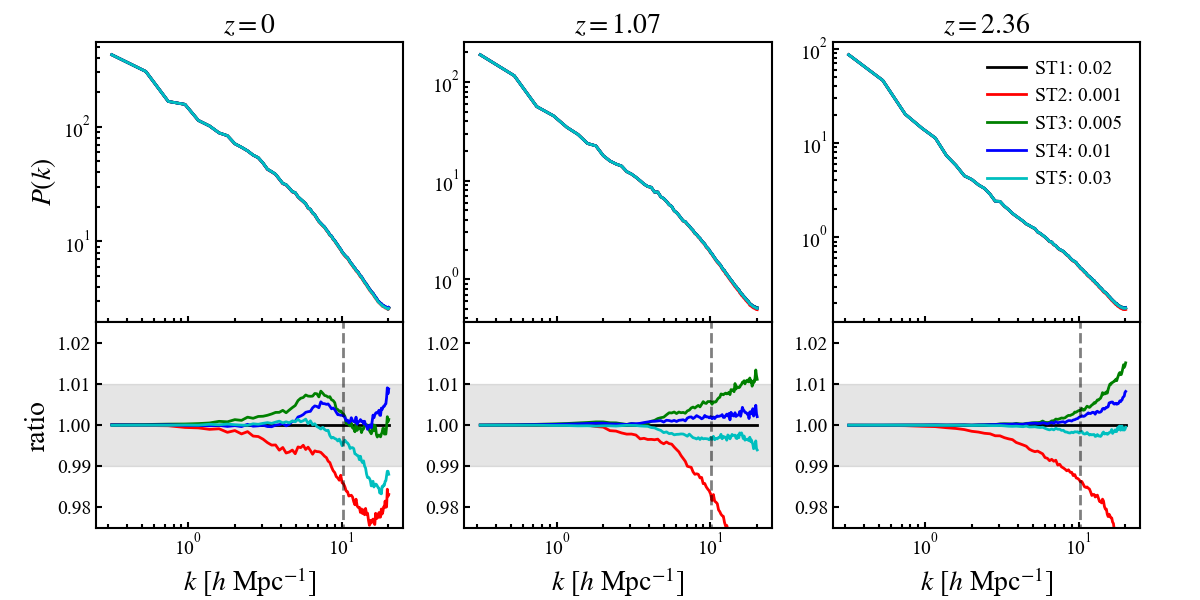}
    \caption{Matter power spectra measured at $z=0$, $z=1.07$, and $z=2.36$ to study the dependence on the softening length. The different coefficients of $\xi=0.001, 0.005, 0.01, 0.02,$ and $0.03$ are shown as red, green, blue, black, and cyan curves, respectively. The lower panel shows the ratio to the fiducial model with $\xi=0.02$; the shaded region indicates a 1\%. The vertical dashed line marks $0.5k_{\rm Nyquist}$.
    }
    \label{fig:softening-1}
\end{figure}

A set of simulations is designed to quantify the impact of the softening length, in which $\xi$ takes values of $0.001, 0.005, 0.01, 0.02, 0.03$, respectively, while the box size and particle number are fixed at $L_{\rm box} = 29.7619\,h^{-1}\,{\rm Mpc}$ and $N_{\rm part}^3 = 192^3$, so that the resolution roughly matches that of the HM simulation. All runs start from identical initial conditions and adopt the same numerical configurations. The resulting matter power spectra are shown in Figure~\ref{fig:softening-1}.

As shown in Figure~\ref{fig:softening-1}, the softening length predominantly affects small-scale cosmic structures, which manifests in the matter power spectrum at wavenumbers $k \gtrsim 1\,h\,{\rm Mpc}^{-1}$. This behavior is consistent with physical intuition: a smaller softening length corresponds to a harder collision core, and hence yields a higher power-spectrum amplitude, in agreement with previous studies. However, for the nearly unsoftened case (\(\xi=0.001\)), the trend is reversed: the power spectrum is suppressed. This suppression might be caused by an overly hard core, which scatters colliding particles to larger scales and thereby smears out the density contrast. In addition, the impact of the softening length weakens toward higher redshift, and the corresponding differences in the power spectra are gradually reduced. In particular, below half the Nyquist wavenumber, $k < \frac{1}{2} k_{\rm Nyquist}$ (gray vertical dashed line), the overall amplitude differences remain within 1\%, taking the fiducial case $\xi=0.02$ as the reference \citep{Colombi_2009, Schneider2011, Smith_2014}.  

\begin{figure}[htpb]
    \centering
    \includegraphics[width=0.95\linewidth]{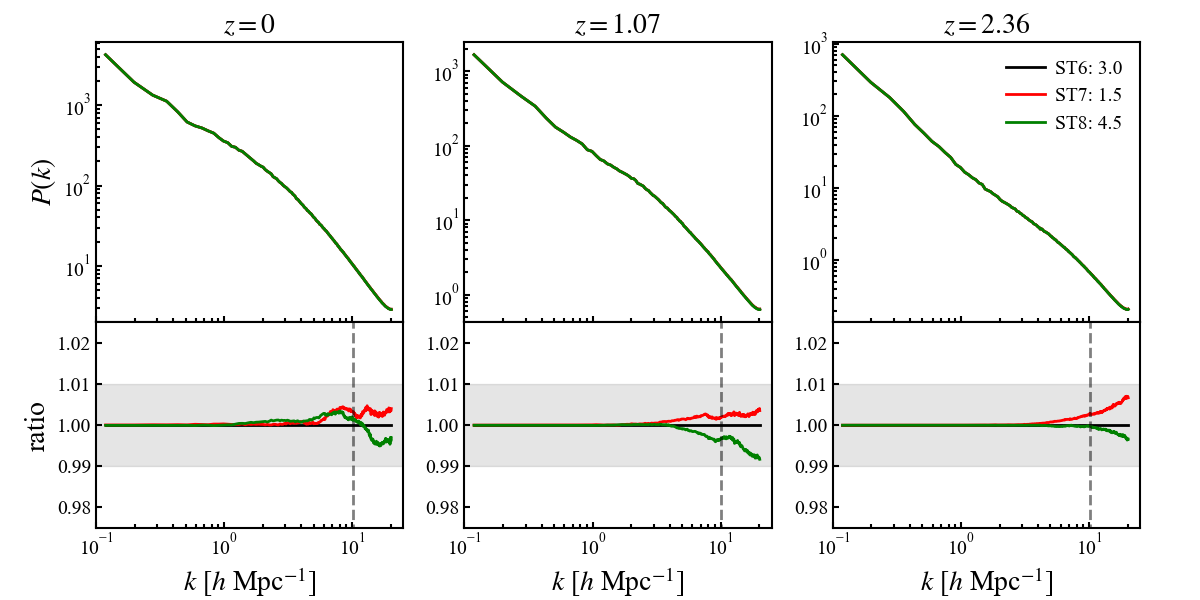}
    \caption{Matter power spectra measured at $z=0$, $z=1.07$, and $z=2.36$ to test convergence with respect to the softening length in the HM simulation. The different values of $\varepsilon_{\rm soft}=1.5, 3.0,$ and $4.5\,h^{-1}\,\mathrm{kpc}$ are shown as red, black, and green curves, respectively. The lower panel shows the ratio to the fiducial model with $\varepsilon_{\rm soft}=3\,h^{-1}\,\mathrm{kpc}$; the shaded region indicates a 1\% deviation. The vertical dashed line marks $0.5k_{\rm Nyquist}$}.
    \label{fig:softening-2}
\end{figure}

We then investigate the softening effect in an enlarged box with the same resolution and softening length in the unit of $h^{-1}\,{\rm kpc}$ as the HM run. We instead perform tests in the box of $L_{\rm box} = 79.3651\,h^{-1}\,{\rm Mpc}$ with $N_{\rm part}^3 = 512^3$ particles. Three softening lengths are considered, $\varepsilon_{\rm soft} = 1.5$, $3.0$, and $4.5\,h^{-1}\,{\rm kpc}$, corresponding approximately to $\xi = 0.01$, $0.02$, and $0.03$, respectively. The resulting matter power spectra are presented in Figure~\ref{fig:softening-2}.

The results present the same qualitative trend as in the previous tests, with smaller deviations at $z = 0$ owing to the reduced cosmic variance. With this additional validation, we find that all three softening choices achieve an accuracy of approximately $0.5\%$ across all statistical scales and at all three redshifts.

\subsection{Adaptive stepping}
\label{sec:adap}

In practical numerical simulations, fixed time steps, although accurate, usually require an unacceptably long computation time. An efficient adaptive time-stepping scheme is therefore an indispensable solution. Meanwhile, adaptive stepping may break the symplectic structure and hence spoil energy conservation, and it may also introduce additional numerical errors \citep{GADGET2, photons, 2020MNRAS.493.1913H}. Since we adopt a long–short range decomposition of the gravitational force in our PM-FMM algorithm, adaptive stepping will only affect small-scale structures. Specifically, we employ the conventional power-of-two block stepping scheme, which is formulated as follows:
\begin{equation}
    \sqrt{\frac{2\eta\varepsilon_{\rm soft} a^3}{G|\mathbf{a}_{i}|}}<\frac{\Delta \ln a}{2^n},
\end{equation}
where $\mathbf{a}_{i}$ is the acceleration of particle $i$, the parameter $\eta = 0.035$ is set in analogy to the GADGET scheme, and $\Delta \ln a$ denotes the base step of the simulation in logarithmic scale factor.

\begin{figure}[htpb]
    \centering    
    \includegraphics[width=0.7\linewidth]{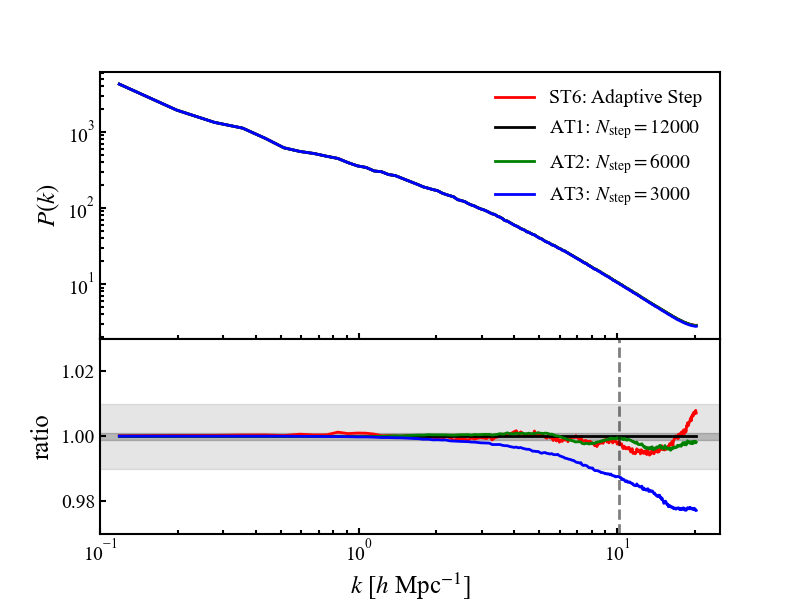}
    \caption{Convergence test for the adaptive time-stepping scheme. The lower panel shows the ratio with respect to $N_{\rm step}= 12000$ fixed-step run. The light and dark shaded regions denote the 1\% and 0.1\% deviation, respectively. The vertical dashed line indicates $0.5k_{\rm Nyquist}$.}
    \label{fig:timestep}
\end{figure}

We also employ a simulation box of $L_{\rm box} = 79.3651\,h^{-1}\,{\rm Mpc}$ with $N_{\rm part}^3 = 512^3$ particles for the time-stepping scheme convergence test. As references, we select three fixed-step runs with total step numbers of $N_{\rm step} = 3000$, $6000$, and $12000$, respectively, For the adaptive run, the base-level step is set to 4000, resulting in approximately 12000 total steps. The resulting matter power spectra are presented in Figure~\ref{fig:timestep}.

We expect the adaptive scheme to produce results consistent with the corresponding fixed-step runs. As shown in Figure~\ref{fig:timestep}, the adaptive run achieves an accuracy of $\sim 0.1\%$ relative to the fixed-step reference with $N_{\rm step} = 12000$ at $k < 0.5 k_{\rm Nyquist}$, and it also shows convergence when compared with the $N_{\rm step} = 6000$ result. The deviation increases to about $0.5\%$ when the full range of scales is considered. We therefore conclude that the adaptive stepping scheme delivers reliable results at the sub-percent accuracy level in our code. For clarity, the numerical simulation parameters used in Section~\ref{sec:soft} and \ref{sec:adap} are summarized in Table \ref{tab:test_sim}.

\begin{table}[htpb]
    \centering
    \begin{tabular}{ccccccccc}
    \hline
    \hline
     & $L_{\rm box}$ [$h^{-1}\,{\rm Mpc}$] & $N_{\rm part}$ & $N_{\rm grid}$ & $\varepsilon_{\rm soft}$ & $\theta$ & $R_{\rm cut}$ & $N^{\rm max}_{\rm reaf}$ & Time-stepping \\
    \hline
    ST1 & 29.7619 & 192 & 192 & $0.02(L_{\rm box}/N_{\rm part})$ & 0.4 & 4.5 & 256 & Adaptive (4000) \\
    ST2 & 29.7619 & 192 & 192 & $0.001(L_{\rm box}/N_{\rm part})$ & 0.4 & 4.5 & 256 & Adaptive (4000) \\
    ST3 & 29.7619 & 192 & 192 & $0.005(L_{\rm box}/N_{\rm part})$ & 0.4 & 4.5 & 256 & Adaptive (4000) \\
    ST4 & 29.7619 & 192 & 192 & $0.01(L_{\rm box}/N_{\rm part})$ & 0.4 & 4.5 & 256 & Adaptive (4000) \\
    ST5 & 29.7619 & 192 & 192 & $0.03(L_{\rm box}/N_{\rm part})$ & 0.4 & 4.5 & 256 & Adaptive (4000) \\
    \hline
    ST6 & 79.3651 & 512 & 512 & $3\,h^{-1}\,{\rm kpc}$ & 0.4 & 4.5 & 256 & Adaptive (4000) \\
    ST7 & 79.3651 & 512 & 512 & $1.5\,h^{-1}\,{\rm kpc}$ & 0.4 & 4.5 & 256 & Adaptive (4000) \\
    ST8 & 79.3651 & 512 & 512 & $4.5\,h^{-1}\,{\rm kpc}$ & 0.4 & 4.5 & 256 & Adaptive (4000) \\
    \hline
    AT1 & 79.3651 & 512 & 512 & $3\,h^{-1}\,{\rm kpc}$ & 0.4 & 4.5 & 256 & Fixed (12000) \\
    AT2 & 79.3651 & 512 & 512 & $3\,h^{-1}\,{\rm kpc}$ & 0.4 & 4.5 & 256 & Fixed (6000) \\
    AT3 & 79.3651 & 512 & 512 & $3\,h^{-1}\,{\rm kpc}$ & 0.4 & 4.5 & 256 & Fixed (3000) \\
    \hline
    \end{tabular}
    \caption{The parameter list of all simulations employed for accuracy and convergence tests.}
    \label{tab:test_sim}
\end{table}

\subsection{Time to solution}

We already have examined how the accuracy responds to changes in the softening scheme and the adaptive stepping configuration, and we have also quantified the speedup brought by the optimizations. At the sub-percent accuracy level required by observations, we now compare the computational time of these simulation runs, as summarized in Table~\ref{tab:test_time}.

\begin{table*}[ht]
    \centering
    \begin{tabular}{ccccccc}
        \hline 
        \hline
     & \multicolumn{3}{c}{Adaptive step} & \multicolumn{3}{c}{Fixed step} \\
     & \multicolumn{3}{c}{$\varepsilon_{\rm soft}$ [$h^{-1}\,{\rm kpc}$]} & \multicolumn{3}{c}{$N_{\rm step}$ [$\varepsilon_{\rm soft}=3\,h^{-1}\,{\rm kpc}$]} \\
    \hline
    Parameter & 1.5 & {\bf 3 (fid)} & 4.5 & 12,000 & 6,000 & 3,000 \\
    \hline
    Time [s] & 122,501 & {\bf 87,619} & 74,013 & 152,139 & 76,709 & 37,562 \\
    \hline
    \end{tabular}
    \caption{Wall-clock time for all $N_{\rm part}=512$ simulations. The fiducial simulation employs an adaptive scheme with $\varepsilon_{\rm soft}=3\,h^{-1}\,{\rm kpc}$, which corresponds to fixed-step run with 12,000 steps. The softening length is set to $3\,h^{-1}\,{\rm kpc}$ in three fixed-step runs.}
    \label{tab:test_time}
\end{table*}

Besides the softening length and adaptive stepping, we have also varied other parameters considered in the algorithm, such as the opening angle $\theta$, $R_{\rm cut}$ and $N_{\rm leaf}^{\rm max}$. We find that their impact on the accuracy is very tiny, typically at the level of $10^{-3.5}$. Moreover, these effects do not accumulate in a single direction: when multiple parameters are varied simultaneously, their contributions to the error largely cancel one another, and the total deviation remains at the level of $10^{-3}$. We therefore restrict our attention to the two key factors, namely the softening length and adaptive stepping.

\section{Code Comparison}

In the previous section, we examined the internal convergence of the PhotoNs-3 code. In this section, we turn to code comparisons against the well-established GADGET code.

A necessary optimization of PhotoNs-3: quantization, is employed in the HM run to reduce the memory limit. To verify that this treatment does not compromise accuracy, we compare PhotoNs-3 against GADGET run in double precision, and examine statistics such as the matter power spectrum and the halo/subhalo mass function. We also pay particular attention to small-scale accuracy, including dark matter halo profiles and positions during the cosmic evolution. Note that we employ the same gravity solver, TreePM, for both Gadget-3 and Gadget-4 in this work, although Gadget-4 additionally provides an FMM-PM implementation.

Therefore, we design four sets of simulations, as listed in Table~\ref{tab:simulation_set}. The fiducial run S1 is carried out with the HM parameters; S2 is an up-scaled counterpart of S1, probing the resolution effect; and S3 and S4 adopt varied cosmologies, with the relevant parameters shifted by $\pm 10\%$. All runs share identical initial conditions.

\begin{table}[ht]
    \centering
    \begin{tabular}{cccccccccc}
    \hline 
    \hline
    & $L_{\rm box}$ & $N_{\rm part}^3$ & $m_p$ & $\Omega_M$ & $\Omega_\Lambda$ & $\sigma_8$ & $H_0$ & $n_s$ & $\varepsilon_{\rm soft}$ \\
    & [$h^{-1}\,{\rm Mpc}$] & & [$h^{-1}\,{\rm M}_\odot$] & & & & [${\rm km}\,{\rm s}^{-1}\,{\rm Mpc}^{-1}$] & & [$h^{-1}\,{\rm kpc}$] \\
    \hline
    S1 & 158.73 & 1024$^3$ & $3.22\times 10^8$ & 0.3111 & 0.6889 & 0.8102 & 67.66 & 0.9665 & 3.0 \\
    S2 & 2000 & 1024$^3$ & $6.43\times 10^{11}$ & 0.3111 & 0.6889 & 0.8102 & 67.66 & 0.9665 & 39.0625 \\
    S3 & 79.3651 & 512$^3$ & $3.54\times 10^8$ & 0.3422 & 0.6578 & 0.8912 & 74.43 & 1.0632 & 3.0 \\
    S4 & 79.3651 & 512$^3$ & $2.89\times 10^8$ & 0.2800 & 0.7200 & 0.7292 & 60.89 & 0.8699 & 3.0 \\
    \hline
    \end{tabular}
    \caption{The simulation suites for code comparison. S1, S3 and S4 shares the same resolution as HM while S2 uses an enlarged box. The S1 and S2 adopt HM parameters, while S3 and S4 introduce a +10\% and −10\% shift of cosmological parameters, respectively. All simulations are carried out in the same initial conditions consistently generated by \texttt{NgenIC} in Gadget-4.}
    \label{tab:simulation_set}
\end{table}

Due to algorithms and implementations, any two codes can hardly reproduce the evolution in perfect agreement \citep{CCP08, DESI22}. To measure the level of differences, we compare our results with both Gadget-3 (In practice, we use a modified version of Gadget-3, referred to as P-Gadget3.) and Gadget-4: the discrepancy between the two GADGET versions provides a useful reference scale for assessing the difference level among software implementations\footnote{Although Gadget-3 has no public release, it has been widely used as a popular version in many influential numerical simulations. In this work, Gadget-4 is used in its publicly released version. For both codes, we made no source-code-level optimizations or modifications.}. For the GADGET runs, the configuration parameters are set as follows: PMGRID = $N_{\rm part}$, ASMTH = 1.25, RCUT = 4.5, ErrTolIntAccuracy = 0.035, MaxSizeTimestep = 0.01, ErrTolTheta = 0.7, ErrTolForceAcc = 0.005, TopNodeFactor = 2.5 and ActivePartFracForDomainDecomp = 0.01. For PhotoNs-3, we adopt the default settings in Section 2 to ensure similar parameters hold the same value as those in the GADGET code. Specifically, $N_{\rm grid}$ (PMGRID) = $N_{\rm part}$, $r_s$ (ASMTH) = 1.25, $R_{\rm cut}$ = 6 and $\eta$ (ErrTolIntAccuracy) = 0.035.

\subsection{Density Field}

We present the pixelized density-field of the S1 and S2 simulations at $z = 0$ \citep{Bayer}. Specifically, we assign particles onto a regular grid using the Nearest-Grid-Point (NGP) scheme. Since the two runs employ boxes of different sizes, we use a slice of $6\,h^{-1}\,{\rm Mpc}$ thickness for S1 and a $20\,h^{-1}\,{\rm Mpc}$ slice for S2 along $Z$ coordinate, both projected onto $1024 \times 1024$ images. 

\begin{figure}[htpb]
    \centering
    \includegraphics[width=1.0\linewidth]{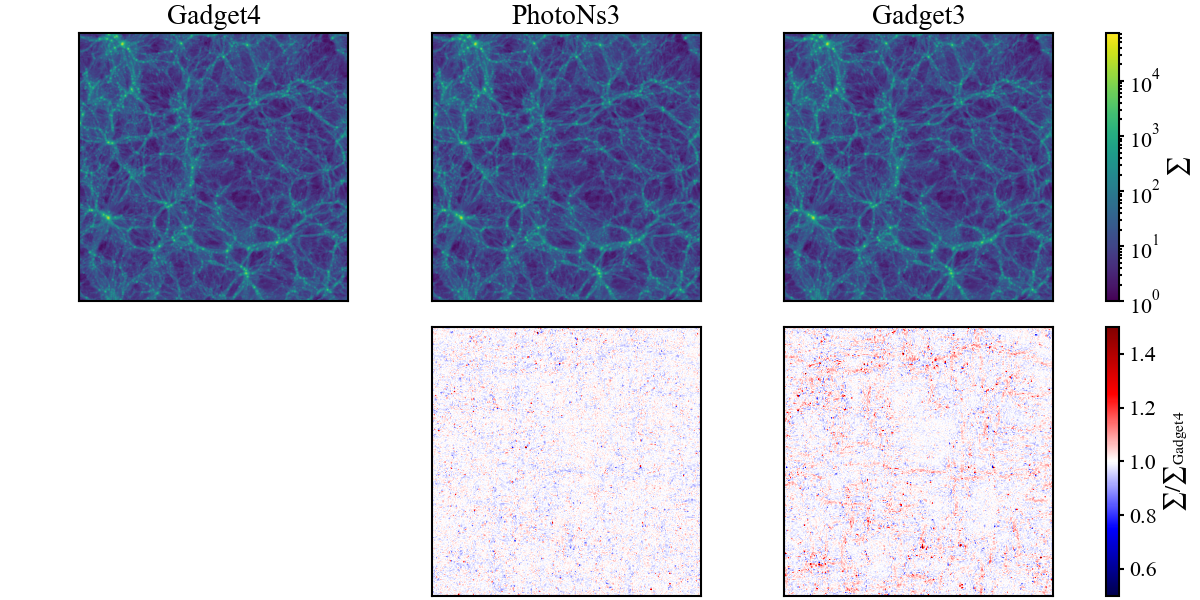}
    \caption{Density field for S1. The upper panel shows DM density slice with a thickness of 6 $h^{-1}\,{\rm Mpc}$. The lower panel shows the density ratio relative to Gadget-4.
    }
    \label{fig:snapshot-S1}
\end{figure}

\begin{figure}[htpb]
    \centering
    \includegraphics[width=1.0\linewidth]{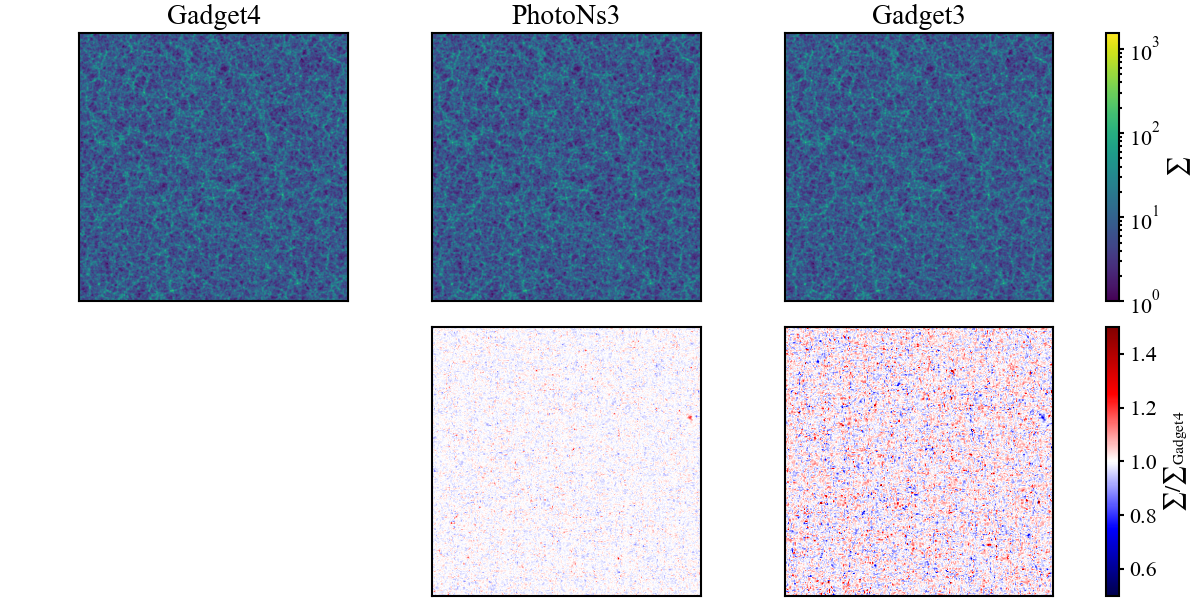}
    \caption{Density field for S2. The upper panel shows DM density slice with a thickness of 20 $h^{-1}\,{\rm Mpc}$. The lower panel shows the density ratio relative to Gadget-4.}
    \label{fig:snapshot-S2}
\end{figure}

Figures~\ref{fig:snapshot-S1} and \ref{fig:snapshot-S2} display the density fields obtained with Gadget-4, PhotoNs-3, and Gadget-3, respectively. All code demonstrate good agreement between each other. To highlight the differences, we take Gadget-4 as the reference and show the relative density ratio $\Sigma/\Sigma_{\rm Gadget4}$ in the lower panels of each figure. It is evident that PhotoNs-3 agrees with Gadget-4 more closely than Gadget-3 does. 

\begin{figure}[htpb]
    \centering
    \includegraphics[width=0.8\linewidth]{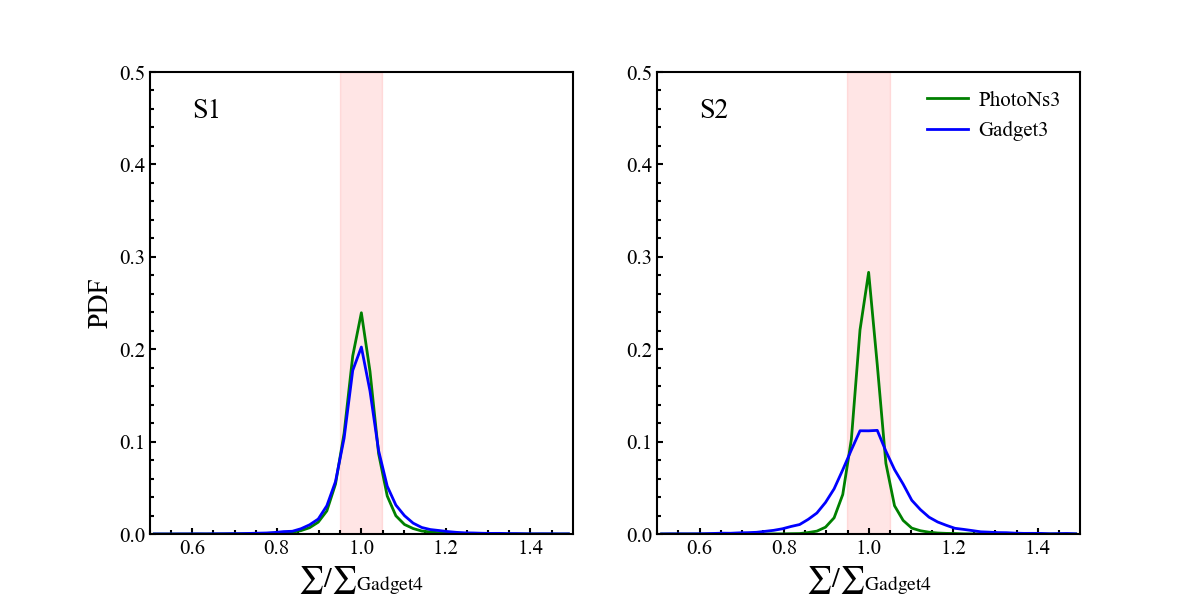}
    \caption{The probability distribution functions of the deviations for PhotoNs-3 and Gadget-3. The red shadow represents the 5\% deviation. The mean values in S1 and S2 are −0.26\% and −0.24\%, respectively, for PhotoNs-3, and 0.19\% and 0.47\%, respectively, for Gadget-3. The corresponding standard deviations are 4.9\% and 3.7\% for PhotoNs-3, and 6.9\% and 9.6\% for Gadget-3.}
    \label{fig:deviation}
\end{figure}

Quantitatively, Figure~\ref{fig:deviation} shows the histogram of the ratio, where the distribution for PhotoNs-3 is more narrowly centered, with the typical deviation from Gadget-4 reduced by about $10\%$ than Gadget-3.

\subsection{Power Spectrum}

\begin{figure}[htpb]
    \centering
    \includegraphics[width=0.95\linewidth]{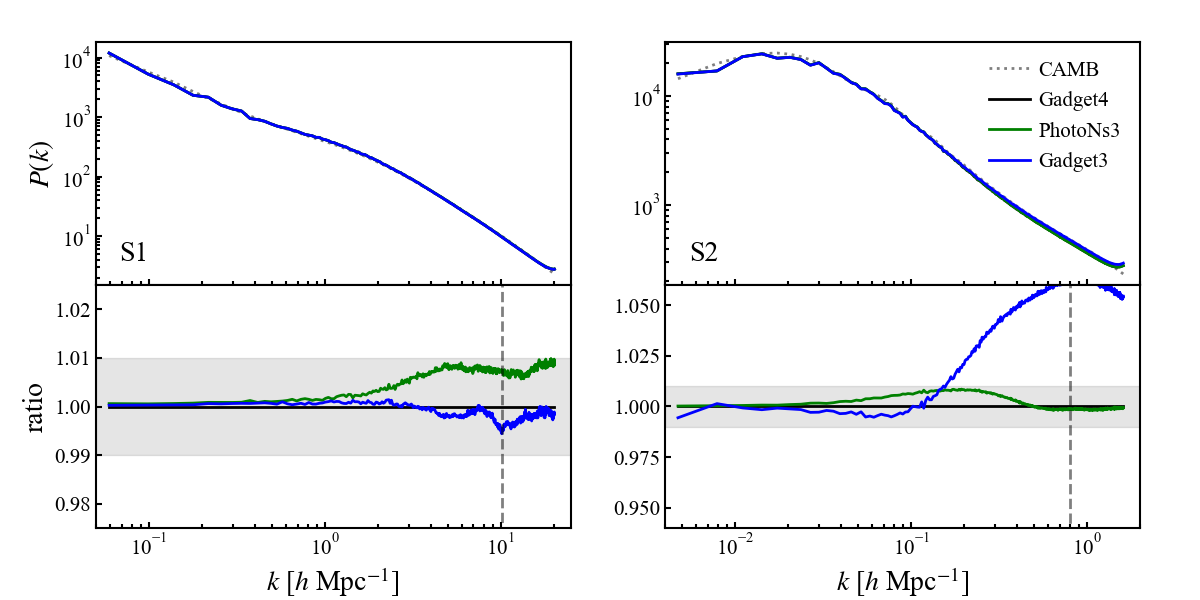}
    \caption{Power spectra for S1 and S2 from the three codes at $z=0$, including a reference of CAMB with Halofit \citep{Halofit} fitting formula. The lower panel shows the ratio relative to Gadget-4. The shaded region indicates the 1\% deviation and the dashed vertical line marks $0.5k_{\rm Nyquist}$.
    }
    \label{fig:power-1024}
\end{figure}

We now turn to the power spectrum comparison and present measurements for all simulation sets with the three codes. Figure~\ref{fig:power-1024} compares the matter power spectra for S1 and S2 at \(z=0\), with GADGET-4 used as the reference in the ratio panels. PhotoNs-3 and GADGET-4 agree to within approximately \(1\%\) over the wavenumber range shown for both simulation sets. The difference between GADGET-3 and GADGET-4 is more pronounced in S2 than in S1: in S2, GADGET-3 yields a higher power-spectrum amplitude on small scales, with the relative difference reaching approximately \(6\%\). This difference can be reproduced under equivalent conditions, so this discrepancy should not be interpreted as a failed run. We speculate that, at the resolution and settings of S2, it may arise from the different numerical convergence trends produced when the same parameters are run in Gadget-3 and Gadget-4. To further examine this difference, we present supplementary comparisons using the Quijote simulation \citep{Quijote} as an external reference in Appendix~\ref{app-quijote}. For both simulation sets, PhotoNs-3 yields slightly higher power than Gadget-4 on nonlinear scales, we partially attribute this to the different softening schemes adopted by the two codes, and a further reference is provided in Appendix~\ref{app-soft}. 

\begin{figure}[htpb]
    \centering
    \includegraphics[width=0.95\linewidth]{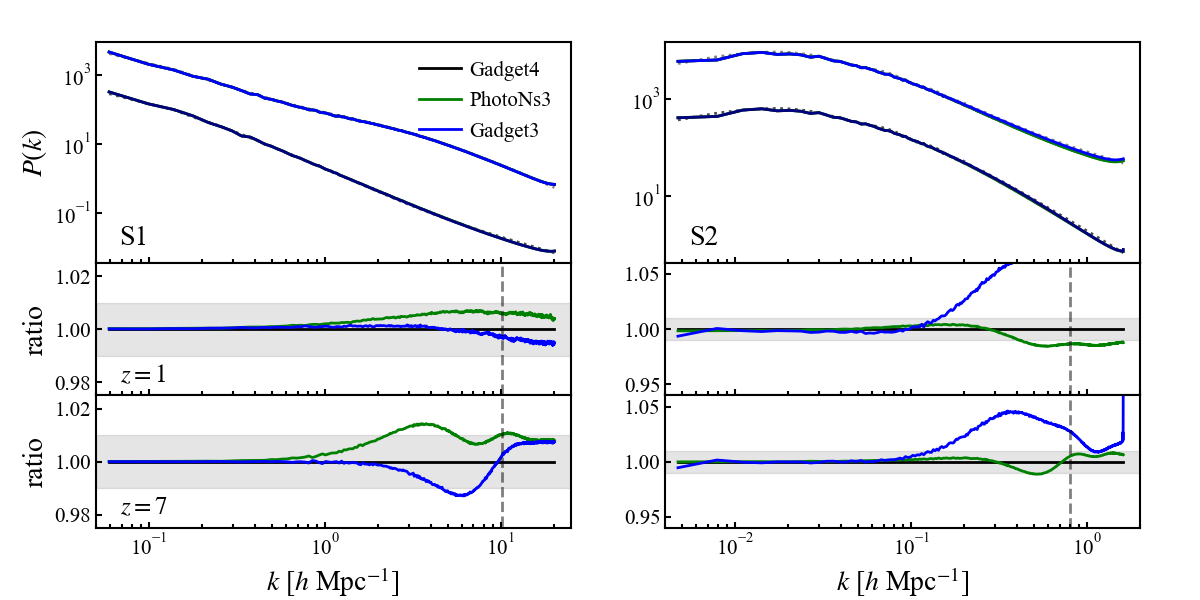}
    \caption{Power spectra for S1 and S2 from the three codes at $z=1$ and $z=7$. The bottom two rows show the ratios relative to Gadget-4. The shaded region indicates the 1\% deviation and the dashed vertical line equals to $0.5k_{\rm Nyquist}$.}
    \label{fig:power-high}
\end{figure}

We also present the S1 and S2 results of the three codes at $z = 1$ and $z = 7$ in Figure~\ref{fig:power-high}. PhotoNs-3 and Gadget-4 maintain excellent consistency throughout the evolution, within roughly sub-percent deviations, despite the different numerical methods employed by the two codes. For S1, the same level of consistency is also found between PhotoNs-3 and Gadget-3, whereas for S2 the Gadget-3 result remains much higher than that of Gadget-4.

\begin{figure}[htpb]
    \centering
    \includegraphics[width=0.95\linewidth]{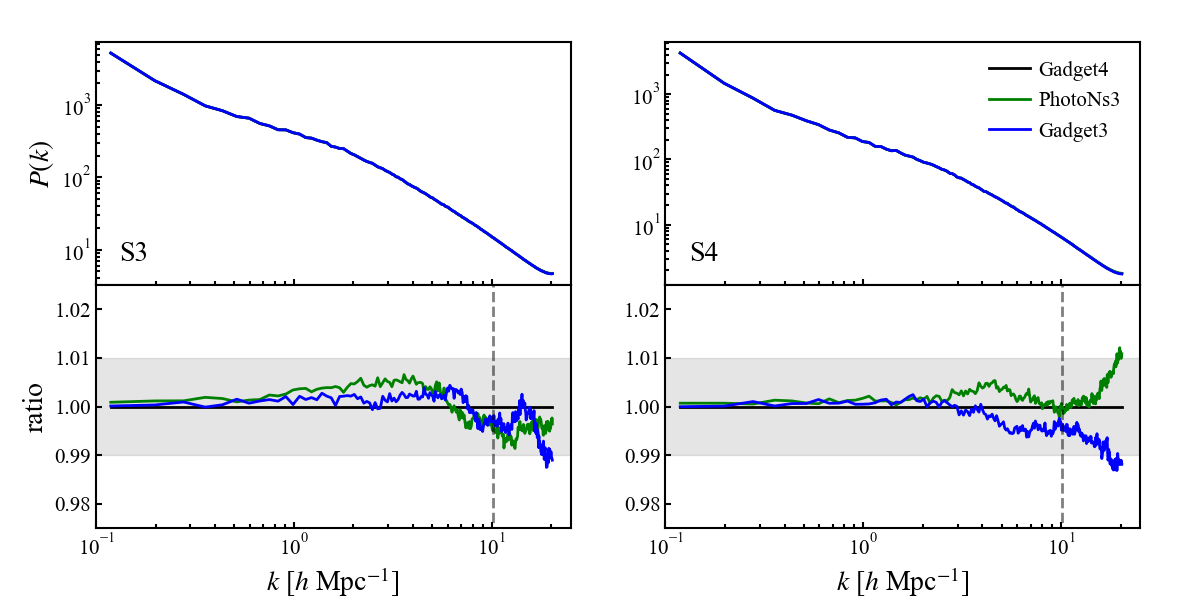}
    \caption{Power spectra for S3 and S4 from the three codes. The lower panel shows the ratio relative to Gadget-4. The shaded region indicates the 1\% deviation and the dashed vertical line equals to $0.5k_{\rm Nyquist}$.}
    \label{fig:power-512}
\end{figure}

Figure~\ref{fig:power-512} presents the power spectra for S3 and S4, which probe the effect of cosmology. The deviations of both PhotoNs-3 and Gadget-3 relative to Gadget-4 remain within $1\%$ across all statistical scales, demonstrating excellent mutual agreement among the codes.

\subsection{Halo Density Profile}

In this subsection, we focus on the density distribution of halos, based on the fiducial simulation (S1).

\begin{figure}[tpb]
    \centering
    \includegraphics[width=0.9\linewidth]{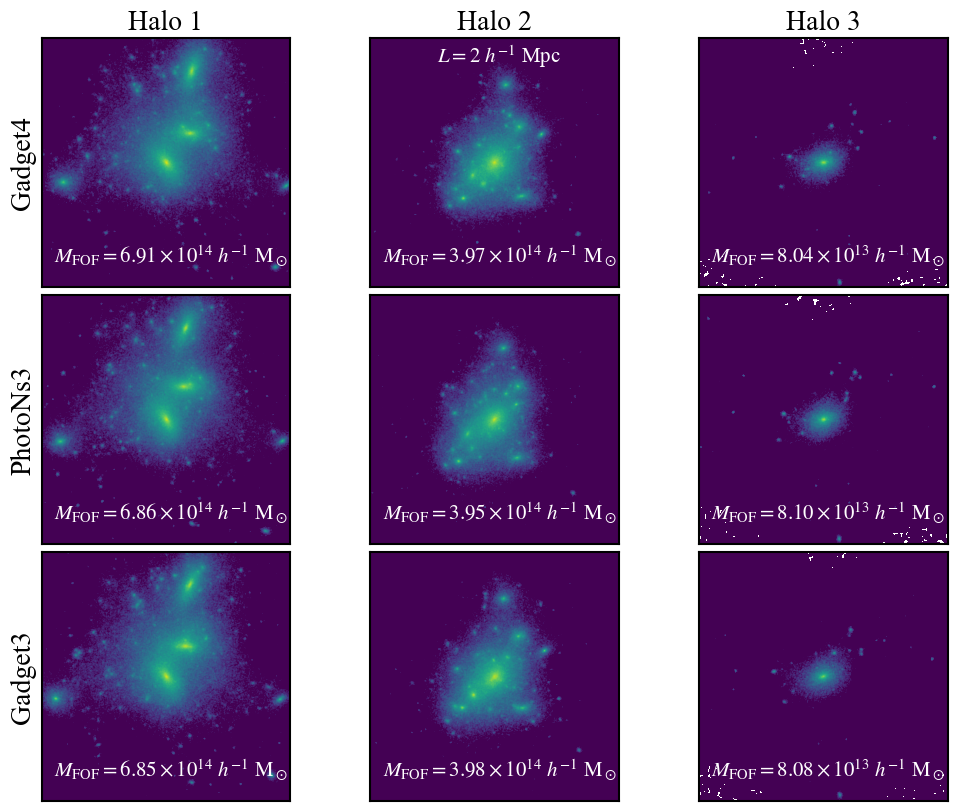}
    \caption{The figure shows the DM density maps in the $X-Y$ plane for the three typical halos. Particle projection are presented in a cubic region of $2\,h^{-1}\,{\rm Mpc}$.}
    \label{fig:halo}
\end{figure}

\begin{figure}[htpb]
    \centering
    \includegraphics[width=1.0\linewidth]{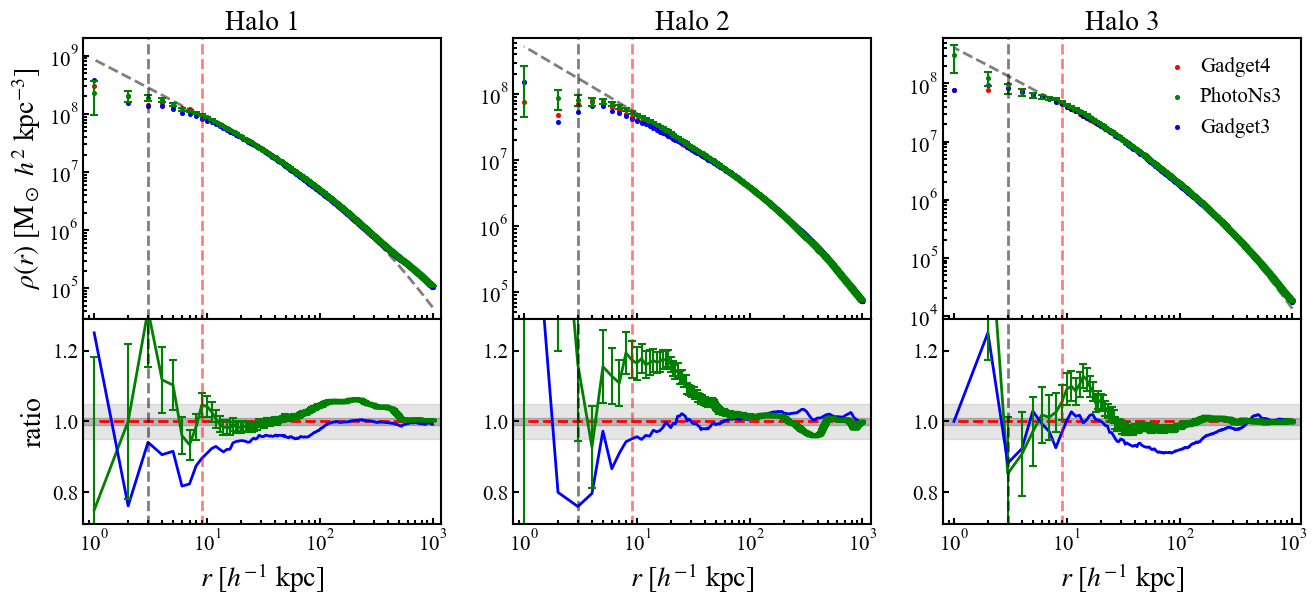}
    \caption{Density profiles of the three typical halos. The upper panel presents the results with NFW fits derived from Gadget-4, the lower panel shows the ratio relative to Gadget-4. The light and dark shaded regions denote the 5\% and 1\% accuracy, the black vertical dashed line indicates $\varepsilon_{\rm soft}$ while the red one indicates $3\,\varepsilon_{\rm soft}$.
    }
    \label{fig:profile}
\end{figure}

Three types of halos are chosen to present in Figure~\ref{fig:halo}, and the corresponding density profiles are shown in Figure~\ref{fig:profile}, respectively. Halo 1 is experiencing a major merger and hosts three massive subhalos; it is also the most massive halo of the three. Its density profile shows a slight overestimation at large radii relative to the NFW curve. Halo 2, a virialized halo which is the second most massive, exhibits a standard profile with a single main subhalo and matches the fitting formula well, especially for PhotoNs-3. Halo 3 is a relatively `dwarf' halo, also with only one main subhalo, showing excellent agreement across all three codes.

Meanwhile, Figure~\ref{fig:halo} shows that all three codes exhibit high consistency in resolving massive subhalos. For smaller subhalos, PhotoNs-3 shows better agreement with those resolved by Gadget-4. Such consistency is not trivial in code comparison studies \citep{Euclid23}.

According to the density profiles in Figure \ref{fig:profile}, we find that the three codes agree very closely at $r > 3\,\varepsilon_{\rm soft}$, although tiny deviations remain. One source of these differences is the distinct softening kernels adopted by the codes, whose effect is most pronounced in the innermost region of the halo. Combined with the projected density maps shown previously, we can confirm that all three codes consistently capture the trajectories and motions of dark matter particles within the virialized region remarkably well.

The three curves in Figure \ref{fig:profile} are fitted with the NFW profile. The best-fitting parameters $(R_s/[10^6~h^{-1}\,{\rm kpc}],\rho_0/[{\rm M}_\odot\,h^2\,{\rm kpc}^{-3}]) $ for Halo 1, Halo 2, and Halo 3 in Gadget-4 run are $(2.84, 304.38)$, $(0.913, 578.14)$, and $(1.91, 218.15)$, respectively.

In addition, we find that the most massive halos identified by the three codes are in one-to-one spatial correspondence and exhibit the same mass ranking. Table~\ref{tab:position} presents the positions of the five most massive halos for the three codes. We further analyze the subhalo (at least 20 DM particles) counts within these five halos: compared to Gadget-4, PhotoNs-3 achieves an average deviation of $0.3\%$, while the deviation relative to Gadget-3 is $2.4\%$.

\begin{table}[ht]
    \centering
    \begin{tabular}{cccc}
    \hline \hline
     &  & Coordinate [$h^{-1}\,{\rm Mpc}$] & \\
         & PhotoNs-3 & Gadget-4 & Gadget-3 \\
    \hline
    Halo 1 & (1.48, 4.06, 156.43) & (1.44, 4.05, 156.46) & (1.46, 4.06, 156.44) \\
    Halo 2 & (25.62, 7.09, 18.00) & (25.59, 7.12, 17.95) & (25.58, 7.11 17.94) \\
    Halo 3 & (1.14, 6.08, 67.38) & (1.12, 6.10, 67.37) & (1.09, 6.10, 67.36) \\
    Halo 4 & (20.53, 11.00, 158.49) & (20.53, 10.98, 158.49) & (20.52, 10.98 158.49) \\
    Halo 5 & (12.53, 2.49, 24.25) & (12.53, 2.48, 24.27) & (12.52, 2.49, 24.27) \\
    \hline
    \end{tabular}
    \caption{The coordinates of the five most massive halos in the S1.}
    \label{tab:position}
\end{table}

\subsection{Mass Function for Halo \& Subhalo }

We now examine both the halo mass function and the subhalo mass function in all simulation sets. PhotoNs-3 employs two subroutines to identify dark matter halos and subhaloes, which provide $M_{\rm FOF}$, $M_{\rm 200c}$, and $M_{\rm 200m}$ for the former, while the latter are identified by a modified \texttt{SUBFIND} algorithm. The catalogs are stored in \texttt{HDF5} format.

\begin{figure}[htpb]
    \centering
    \includegraphics[width=0.95\linewidth]{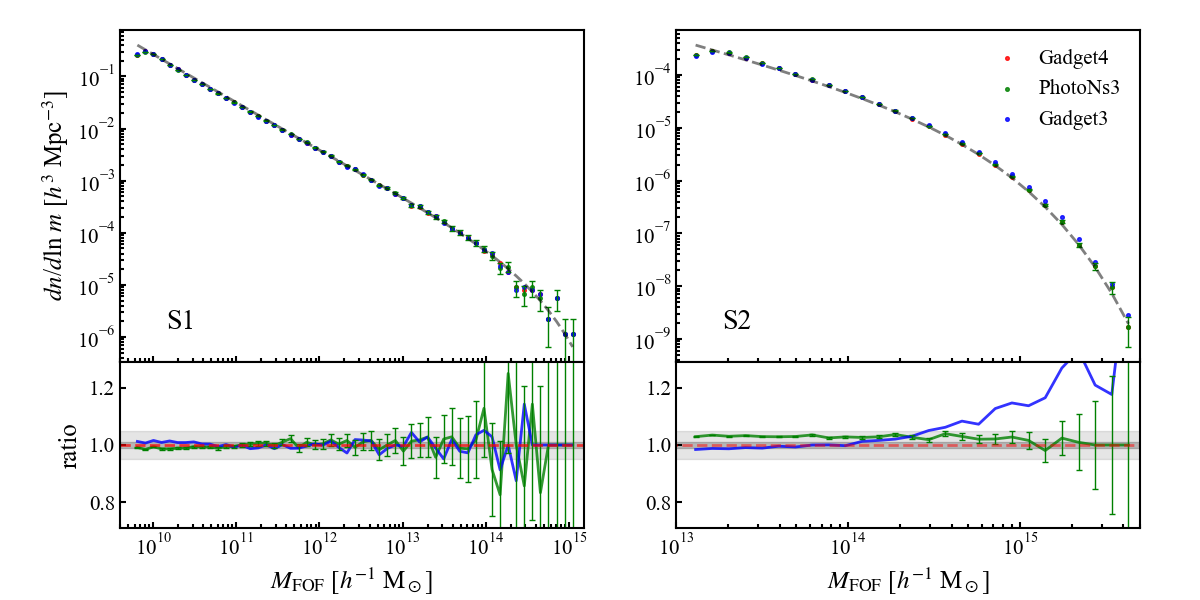}
    \caption{The upper panel presents halo mass functions in S1 and S2 for three codes and the lower panel shows the relative deviations with respect to Gadget-4. The light and dark shaded regions denote the 5\% and 1\% deviation. The fitting function of \citet{watson13} is also shown as a reference.}
    \label{fig:mass-1024}
\end{figure}

Figure~\ref{fig:mass-1024} presents the halo mass function results of S1 and S2 at $z=0$. In S1, both PhotoNs-3 and Gadget-3 agree with Gadget-4 to within $1\%$ for low-mass halos, while the deviations become larger in the high-mass range, accompanied by increasing statistical errors, though still about $5\%$. Note that part of the deviation in the high-mass range is induced by the binning effect. In S2, the halo mass function from PhotoNs-3 also shows good agreement (within $5\%$) with that from Gadget-4. For Gadget-3, the results indicate an overproduction of massive halos relative to Gadget-4, a trend that is also reflected in the power spectrum result.

\begin{figure}[htpb]
    \centering
    \includegraphics[width= 0.95\linewidth]{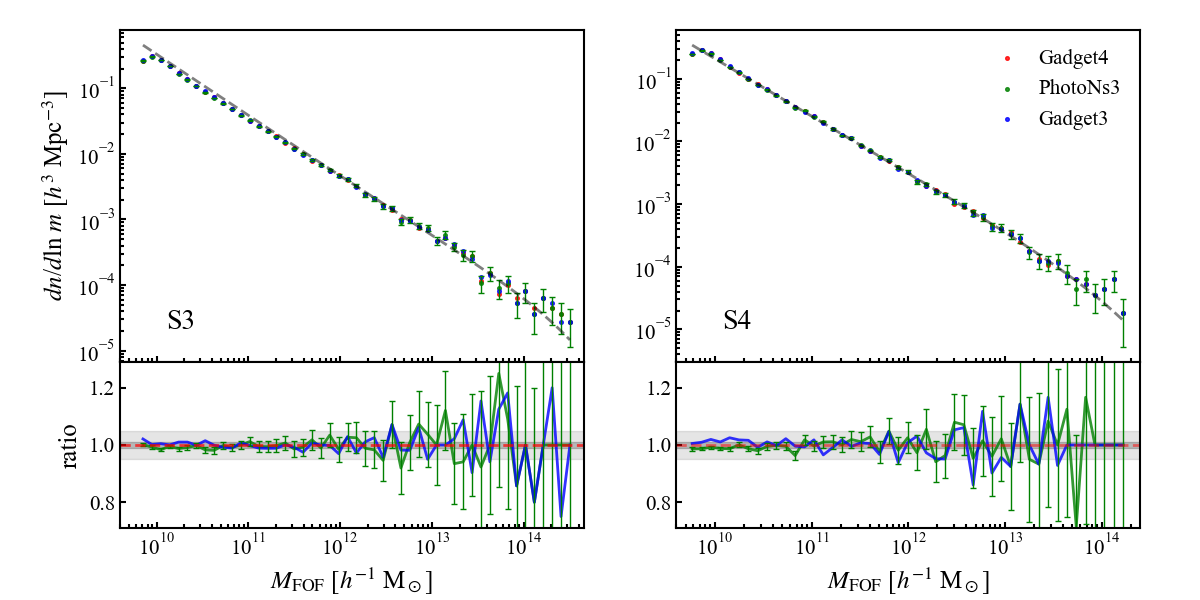}
    \caption{Halo mass functions from S3 and S4 for three codes.}
    \label{fig:mass-512}
\end{figure}

Figure~\ref{fig:mass-512} presents the halo mass function results of S3 and S4, which evolves in different cosmological parameters. We also find a high level of consistency among the three codes. Quantitatively, the halo mass functions of small halos agree to about $1\%$ across the codes in S3, while the deviation increases slightly in S4 but remains within $5\%$.

\begin{figure}[htpb]
    \centering
    \includegraphics[width=1.0\linewidth]{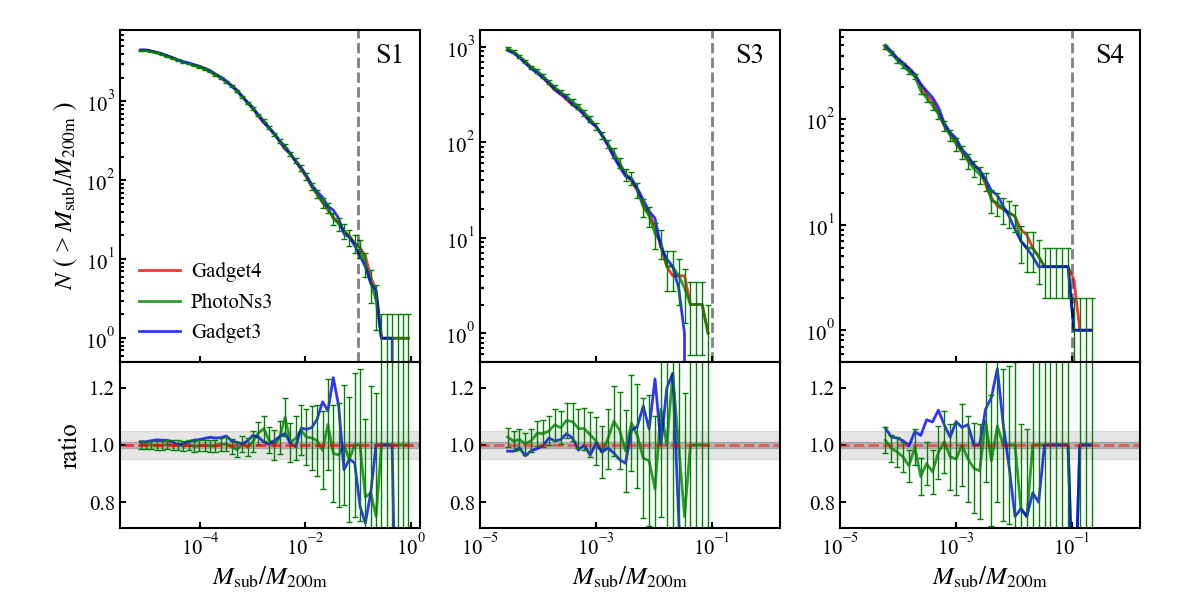}
    \caption{Subhalo mass functions for halos with host halo mass range of $>10^{13}~h^{-1}\,{\rm M}_{\odot}$ in S1, S3 and S4.}
    \label{fig:sub-512}
\end{figure}

Figure~\ref{fig:sub-512} shows the subhalo mass function results. The deviations from Gadget-4 for both PhotoNs-3 and Gadget-3 are slightly larger than those in the halo mass function, roughly within $5\%$ for low-mass subhalos. In general, the subhalo mass functions are highly consistent among the three codes, with deviations at the level of reasonable statistical fluctuations \citep{2004MNRAS.355..819G, 2010MNRAS.404..502G}.

\subsection{Performance}

\begin{figure}[htpb]
    \centering
    \includegraphics[width=0.75\linewidth]{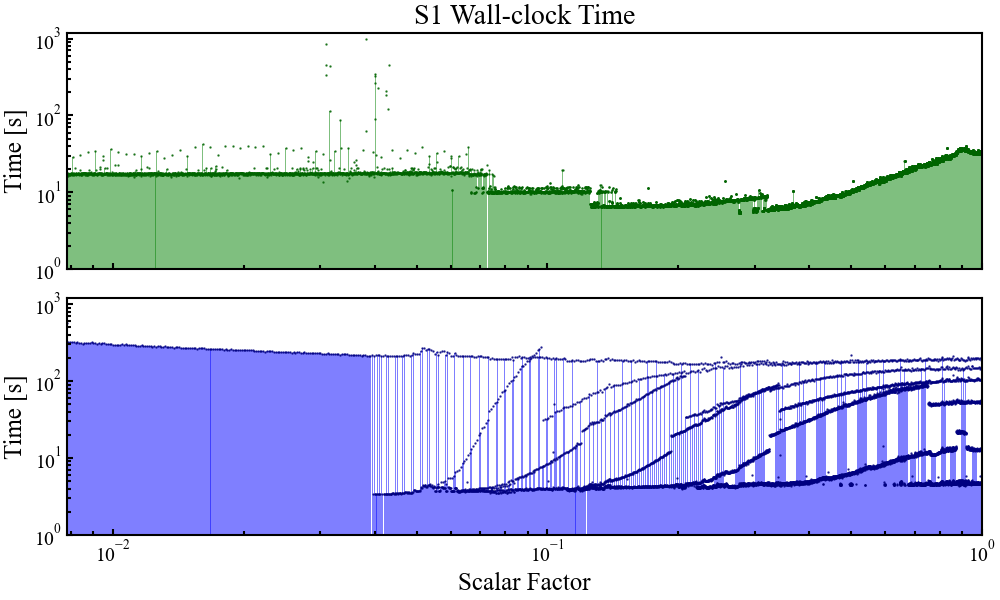}
    \caption{Wall-clock time per time step as a function of the scale factor $a$ in S1. The upper and lower panels show the results from PhotoNs-3 and Gadget-4, respectively. The total wall-clock time for PhotoNs-3 is about 54.1 hours (194752 seconds) and for Gadget-4 is about 90.2 hours (324771 seconds).}
    \label{fig:time-S1}
\end{figure}

\begin{figure}[htpb]
    \centering
    \includegraphics[width=0.75\linewidth]{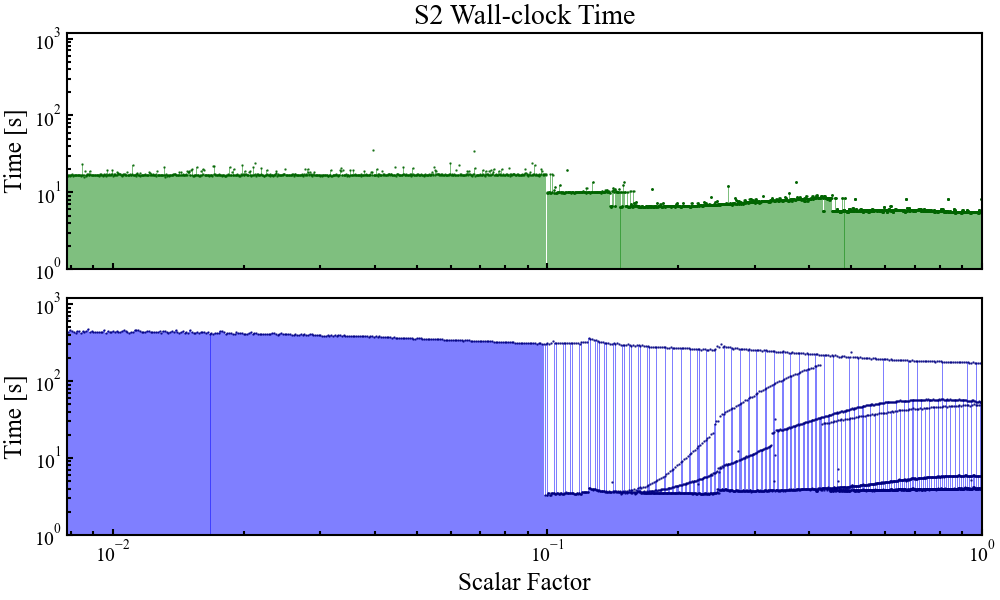}
    \caption{Wall-clock time per time step as a function of the scale factor $a$ in S2. The total wall-clock time for PhotoNs-3 is about 10.5 hours (37657 seconds) and for Gadget-4 is about 53.5 hours (192754 seconds).}
    \label{fig:time-S2}
\end{figure}

In this section, we present the computational performance of PhotoNs-3. The wall-clock time per time step is presented along the cosmic scale factor in Figure \ref{fig:time-S1} and \ref{fig:time-S2}, for S1 and S2, respectively.

It is evident that PhotoNs-3 is a GPU-accelerated, high-performance code, and it is therefore unsurprising that it runs faster than the public version of Gadget-4. The speedup, however, varies considerably with resolution and particle number. Combined with the results from other runs, such as the HM simulation, the speedup of PhotoNs-3 becomes even more pronounced at extreme scales. For the current S1 and S2 simulations at their respective scales and resolutions, we achieve about $2 \sim 5$ times speedup relative to the Gadget-4 code, and about $\sim 20$ times in an individual step. 

In practice, performance is sensitive to various factors, such as hardware and system environment, so the comparison here serves only as a basic reference on ORISE supercomputer.

\section{Conclusion}

In this work, we present a systematic assessment of the accuracy, convergence, and cross-code consistency of PhotoNs-3, employed for the HM simulation. Because this version introduces dedicated optimization techniques and GPU acceleration aimed at extreme-scale simulation runs, quantifying its numerical reliability is a prerequisite for interpreting simulations such as the Ultramarine and HM runs. 

We performed internal convergence tests, examining the dependence of the results on the softening length, the adaptive time-stepping scheme, and remaining algorithmic parameters. We find that PhotoNs-3 achieves well-controlled convergence in all of these aspects. The adaptive time-stepping scheme reproduces the fixed-step reference with $N_{\rm step} = 12000$ to about $0.1\%$ at $k < 0.5 k_{\rm Nyquist}$, with the deviation increasing only to $0.5\%$ over the full range of scales; and the remaining parameters, such as the opening angle, the cutoff radius and the maximum leaf size, affect the accuracy at the level of $10^{-3}$.

We then compared our code against Gadget-3 and Gadget-4 using four simulation sets with identical initial conditions: the fiducial run S1 adopting the HM parameters, an up-scaled run S2 probing the resolution effect, and the other two runs S3 and S4 with the cosmological parameters shifted by $\pm 10\%$. PhotoNs-3 agrees with Gadget-4 at sub-percent level in the matter power spectrum over the statistical scales, redshifts, and cosmologies explored, and the two codes also recover the most massive halos with nearly identical ranking and positions. The consistency also extends to nonlinear observables: the halo mass functions agree to within $1\%$ for low-mass halos and remain within $5\%$ at the high-mass end; the subhalo mass functions are consistent among all three codes within reasonable statistical fluctuations; and the halo density profiles agree closely for $r > 3\,\varepsilon_{\rm soft}$. Notably, in the projected density fields, the residuals of PhotoNs-3 relative to Gadget-4 are about $10\%$ smaller than those of Gadget-3, and the subhalo counts within the five most massive halos deviate from Gadget-4 by only $0.3\%$ on average, compared with $2.4\%$ for Gadget-3. These comparisons show that PhotoNs-3 is not only consistent with, but in several respects closer to Gadget-4 than the two GADGET versions are to each other, and they validate the quantization scheme used in the HM run.

Overall, our tests establish that PhotoNs-3 achieves excellent numerical convergence, maintain remarkable consistency with the GADGET code across a broad range of resolutions, redshifts, and cosmological parameters, and offers substantially higher computational efficiency at the same time. This combination of accuracy and performance makes PhotoNs-3, and the HM simulation in particular, a reliable theoretical basis for precision cosmology with current and forthcoming wide-field surveys.

\normalem
\begin{acknowledgements}
This work is supported by the National Key Research and Development Program of China (Grant No. 2023YFB3002501) and the National Natural Science Foundation of China (NSFC) (Nos. 12588202, 12473002, 12473015).
ML also acknowledges the support from the National SKA Program of China (Nos. 2022SKA0110201) and the National Key Research and Development Program of China (No. 2022YFA1602903). 
The numerical calculations in this study were carried out on the ORISE Supercomputer.
\end{acknowledgements}

\bibliographystyle{raa}      

\appendix                  

\section{Additional Power Spectrum Comparisons at the S2 Resolution}
\label{app-quijote}

To investigate differences between Gadget-3 and Gadget-4 in S2, we introduce Quijote simulations for additional comparisons as an external reference \citep{Quijote}. We carried out five supplementary simulations with the initial condition from one of the Quijote realization (No. 10) \footnote{https://quijote-simulations.readthedocs.io/en/latest/index.html}, using $N_{\rm part}^3=512^3$ particles in a box of $L_{\rm box} = 1000\,h^{-1}\,{\rm Mpc}$ (approximately the same mass resolution of our S2 run).

The configurations differ in, among other settings, the time-integration accuracy parameter \texttt{ErrTolIntAccuracy}, and the PM mesh size \texttt{PMGRID}. In these supplementary tests, the S2 configuration uses $\texttt{ErrTolIntAccuracy}=0.035$ and $\texttt{PMGRID}=512$, while the Quijote configuration uses $\texttt{ErrTolIntAccuracy}=0.025$ and $\texttt{PMGRID}=1024$. Thus, differences in time integration and in the discretization and splitting of gravitational forces may contribute to the measured changes in the power spectra.

\begin{figure}[ht]
    \centering
    \includegraphics[width=0.8\linewidth]{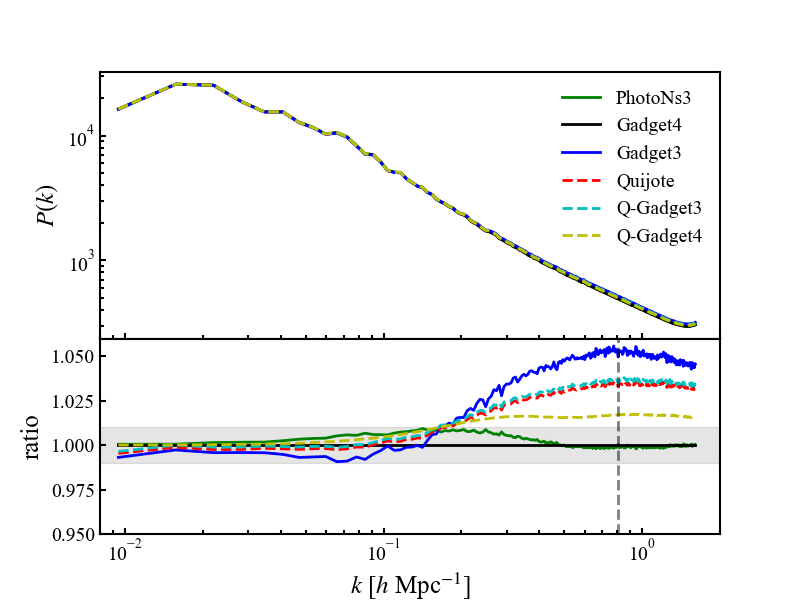}
    \caption{Matter power spectra at $z=0$ for the supplementary comparisons at the S2 resolution. Solid curves show PhotoNs-3 (green), Gadget-4 (black), and Gadget-3 (blue) with the S2 numerical configuration. Dashed curves show the Quijote reference (red), Q-Gadget3 (cyan), and Q-Gadget4 (yellow); the latter two use the Quijote configuration. The lower panel shows ratios relative to the solid black Gadget-4 curve. The shaded band marks a $1\%$ difference from this reference, and the vertical dashed line marks $0.5k_{\rm Nyquist}$.}
    \label{fig:Quijote}
\end{figure}

Figure~\ref{fig:Quijote} compares the matter power spectra at $z=0$. The solid curves show supplementary runs using the S2 configuration, while the dashed curves labelled Q-Gadget3 and Q-Gadget4 use the Quijote configuration. The Quijote spectrum is included as a comparison reference (the red dashed curve).  Comparing the red and cyan curves in the figure, the \textsc{Gadget}-3 run in this work is highly consistent with the external Quijote results. Under the S2 configuration, the rerun \textsc{Gadget}-3 and \textsc{Gadget}-4 simulations reproduce the \(\sim 5\%\) discrepancy reported in Fig.~\ref{fig:power-1024}; when the Quijote configuration are adopted instead, the small-scale discrepancy shrinks from about \(5\%\) to roughly \(2\%\). The similar deviation induced by different PMGRID is also observed in \citet{Smith_2014}. 

Comparisons based on the Quijote simulations do not fully explain the differences between \textsc{Gadget} versions. The dependence of accuracy and convergence on simulation parameters and setup choices is complex, and for different scientific goals the impact of these choices should be examined more carefully.

\section{Complement for Softening Algorithm}
\label{app-soft}

We present the power spectrum results of the softening convergence test carried out by Gadget-4. The test adopts the same simulation setup as the PhotoNs-3 runs ($N_{\rm part}^3 = 512^3$ particles) and identical initial conditions, with three softening lengths $\varepsilon_{\rm soft} = 1.5$, $3.0$, and $4.5\,h^{-1}\,{\rm kpc}$ considered.

\begin{figure}[ht]
    \centering
    \includegraphics[width=0.95\linewidth]{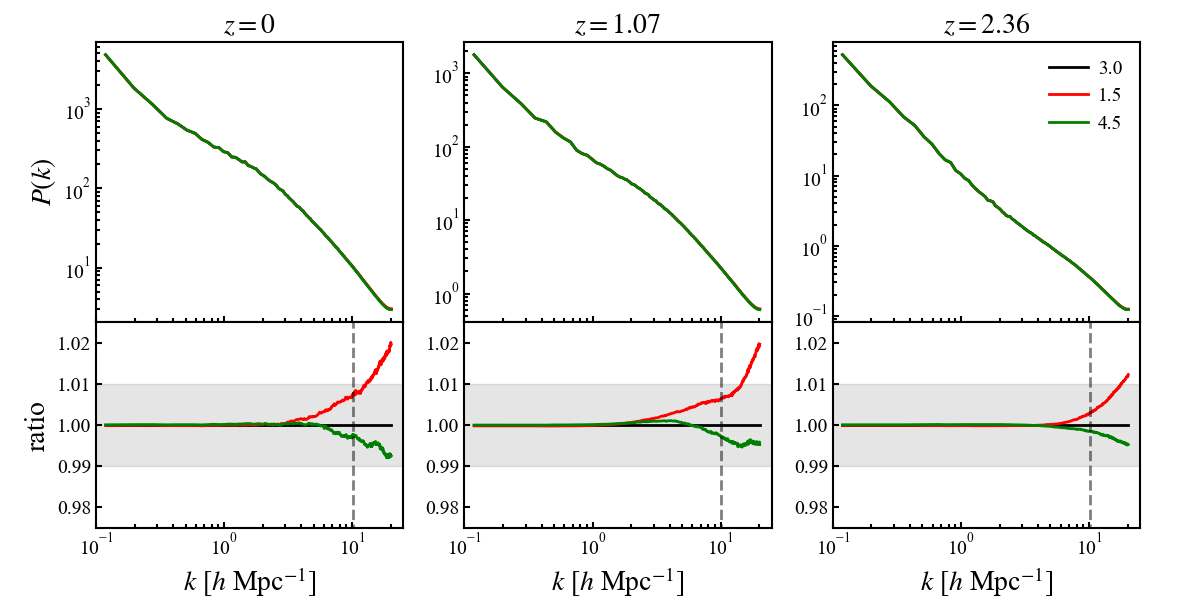}
    \caption{Matter power spectra measured at $z=0$, $z=1.07$ and $z=2.36$ for different $\varepsilon_{\rm soft}$. The lower panel shows the ratio to the fiducial one with $\varepsilon_{\rm soft}=3\,h^{-1}\,{\rm kpc}$ and the shaded region indicating the 1\% deviation. The value of the vertical dashed line is $0.5k_{\rm Nyquist}$.}
    \label{fig:soft-Gadget}
\end{figure}

Figure~\ref{fig:soft-Gadget} reveals a trend similar to that observed in PhotoNs-3, but with larger discrepancies. At $z = 0$, the deviation reaches approximately $1\%$ within $k < 0.5 k_{\rm Nyquist}$ and up to $2\%$ across all statistical scales; this discrepancy persists even at high redshift.

\label{lastpage}

\end{document}